\documentclass[aps, prb,preprint, groupedaddress, floatfix]{revtex4}
\usepackage{epsfig}
\usepackage{graphicx}
\usepackage{amsmath}
\usepackage{bbold}
\usepackage{tabularx, booktabs}
\usepackage[normalem]{ulem}
\usepackage{xcolor}

\newcommand{\etal}{{\it{et al.}}}
\newcommand{\ie}{{\it i.e.}}
\newcommand{\eg}{{\it{e.g.}}}
\newcolumntype{Y}{>{\centering\arraybackslash}X}
\newcolumntype{Z}{>{\hsize=1.1\hsize\centering\arraybackslash}X}

\newcommand*{\PCOO}{Pb$_2$CoOsO$_6$}

\newcommand*{\LOO}{LiOsO$_3$}

\begin{document}

\title{Coupled magnetic and structural phase transitions in
  antiferromagnetic polar metal Pb$_2$CoOsO$_6$ under pressure}

    \author{Yuanyuan Jiao$^{1,2,\&,\P}$}
    \author{Yue-Wen Fang$^{3,4,\P}$}
    \email{fyuewen@gmail.com}
    \author{Jianping Sun$^{1,2}$}
    \author{Pengfei Shan$^{1,2}$} 
    \author{Zhenhai Yu$^{5}$}
    \author{Hai L. Feng$^{1}$}
    \author{Bosen Wang$^{1,2,6}$}
    \author{Hanming Ma$^{7}$}
    \author{Yoshiya Uwatoko$^{7}$}
    \author{Kazunari Yamaura$^{8,9}$}
    \author{Yanfeng Guo$^{5}$}
    \email{guoyf@shanghaitech.edu.cn}
    \author{Hanghui Chen$^{4,10}$}
    \email{hanghui.chen@nyu.edu}
    \author{Jinguang Cheng$^{1,2,6}$}
    \email{jgcheng@iphy.ac.cn}
    \affiliation{
        $^1$Beijing National Laboratory for Condensed Matter Physics and Institute of Physics, Chinese Academy of Sciences, Beijing 100190, China \\
        $^2$School of Physical Sciences, University of Chinese Academy of Sciences, Beijing 100190, China  \\
        $^3$Laboratory for Materials and Structures \& Tokyo Tech World Research Hub Initiative (WRHI), Institute of Innovative Research, Tokyo Institute of Technology, 4259 Nagatsuta, Midori-ku, Yokohama, Kanagawa 226-8503, Japan \\
        $^4$NYU-ECNU Institute of Physics, New York University Shanghai, Shanghai, 200062 China \\
        $^5$School of Physical Science and Technology, ShanghaiTech University, Shanghai, 201210, China \\
        $^6$Songshan Lake Materials Laboratory, Dongguan, Guangdong 523808, China \\
        $^7$Institute for Solid State Physics, University of Tokyo, Kashiwa, Chiba 277-8581, Japan \\
	$^8$International Center for Materials Nanoarchitectonics (WPI-MANA), National Institute for Materials Science, Tsukuba 305-0044, Japan \\
        $^9$Graduate School of Chemical Science and Engineering, Hokkaido University, Sapporo 060-0810, Japan \\
	$^{10}$Department of Physics, New York University, New York, 10012, USA  \\
   $^{\&}$ Present address: Faculty of Science, Wuhan University of Science and Technology, Wuhan, Hubei 430065, China \\
   $^{\P}$ These two authors contributed equally to this work.
    }

\begin{abstract}
\PCOO~is a newly synthesized polar metal in which inversion symmetry
is broken by the magnetic frustration in an antiferromagnetic ordering
of Co and Os sublattices. The coupled magnetic and structural
transition occurs at 45 K at ambient pressure. Here we perform transport
measurements and first-principles calculations to study the pressure
effects on the magnetic/structural coupled transition of \PCOO.
Experimentally we monitor the resistivity anomaly at $T_N$ under
various pressures up to 11 GPa in a cubic anvil cell apparatus. We
find that $T_N$ determined from the resistivity anomaly first
increases quickly with pressure in a large slope of $dT_N/dP$ =
+6.8(8) K/GPa for  $P < 4$ GPa, and then increases with a
much reduced slope of 1.8(4) K/GPa above 4 GPa. Our first-principles
calculations suggest that the observed discontinuity of $dT_N/dP$ around 4 GPa
may be attributed to the vanishing of Os magnetic
moment under pressure. Pressure substantially reduces the Os moment
and completely suppresses it above a critical value, which relieves
the magnetic frustration in the antiferromagnetic ordering of \PCOO.
The Co and Os polar distortions decrease with the increasing pressure
and simultaneously vanish at the critical pressure.
Therefore above the critical
pressure a new centrosymmetric antiferromagnetic state emerges in
\PCOO, distinct from the one under ambient pressure,
thus showing a discontinuity in $dT_N/dP$.
\end{abstract}
\maketitle


\section{Introduction}

Ferroelectricity is usually incompatible with long-range magnetic order
and metallicity~\cite{2006Nature-JFScott,Resta2002JPCM,PhysRevLett.14.217}. Insulating materials which
possess both electric polarization and magnetization
and possibly a coupling between the two order parameters are called
multiferroics, which has been an active research field in condensed matter
physics and materials science~\cite{fang2015first,spaldin2019advances,DongShuaiNSR2019}.
On the other hand, materials with
both polar displacements and intrinsic metallicity are termed as ``polar metals''~\cite{PhysRevLett.14.217,Shi2013},
which have  
attracted increasing interest
in experiment and 
theory~\cite{xiang2015prb,puggioni2014designing,Fei2018,PhysRevMaterials.2.125004,Du-APLMater2019,PhysRevB.99.195154,PhysRevB.101.220101,PhysRevLett.124.237601,PhysRevLett.122.227601,fang2020design}.
Accordering to Anderson and Blount~\cite{PhysRevLett.14.217}, polar metals are characterized by
a second-order structural phase transition with the appearance of
a polar axis and the loss of inversion symmetry at a finite temperature.
LiOsO$_3$ is the first unambiguous example of bulk polar metals, which
transforms from a centrosymmetric $R\bar{3}c$ structure to a polar $R3c$
structure at $T_s=140$ K~\cite{Shi2013,paredes2018pressure}. However, a \textit{metal} that possesses
both polar displacements and magnetization and most importantly
a coupling between the two is extremely rare. Previous theoretical
proposals of cation-ordered SrCaRu$_2$O$_6$~\cite{puggioni2014designing} and BiPbTi$_2$O$_6$~\cite{fang2020design} show
the coexistence of ferromagnetism with polar displacements but
there is no strong coupling between the two properties.

Recently, Princep and coworkers have synthesized a new polar metal
\PCOO~in which an antiferromagnetic order with magnetic frustration
breaks the inversion symmetry~\cite{princep2019magnetically}.  \PCOO~
crystallizes in a cation-ordered double-perovskite structure.  At room
temperature, the material is paramagnetic and centrosymmetric
with space group $P2_1/n$ (No. 14). At $T=45$ K,
\PCOO~undergoes a continuous phase transition to an
antiferromagnetically ordered state with propagation vector
$\textbf{k} = (1/2, 0, 1/2)$. Both Co and Os atoms have magnetic
moments and they order simultaneously.  Accompanying the appearance of
magnetic order is the removal of inversion symmetry. This is because
each Os moment is surrounded by six neighboring Co moments, three of
which are parallel to the Os moment and the other three of which are
antiparallel to the Os moment. The ferromagnetic and antiferromagnetic
couplings between Os and Co moments are generically different and thus
this magnetic frustration forces Os moment to move away from the
centrosymmetric position, which breaks the inversion symmetry. The
low-temperature polar structure has space group $P_c$ (No. 7) with the
corresponding Shubnikov group for the magnetic structure being
$P_ac$. 
Since in \PCOO~the polar displacements are induced by the unique long-range magnetic order,
similar to ``type-II multiferroics''~\cite{type-II2008}, we
may term \PCOO~as ``type-II polar metals'', while SrCaRu$_2$O$_6$~\cite{puggioni2014designing} and
BiPbTi$_2$O$_6$~\cite{fang2020design} may be referred to as ``type-I polar metals'' since
magnetic order and polar displacements 
have different sources and are weakly coupled in these materials. 

In comparison with ferroelectrics, polar metals usually exhibit a much
lower structural transition temperature, \eg~$T_s$ = 140 K for
\LOO~\cite{Shi2013} and $T_s$ (= $T_N$) = 45 K for \PCOO~while the
$T_s$ for a prototypical ferroelectric material BaTiO$_3$ is above
room temperature.  From the viewpoint of practical applications, it is
desirable to increase the transition temperature of polar metals.  By
monitoring the resistivity anomaly at $T_s$ of \LOO~under different
pressures up to 6.5 GPa, Aulestia and coworkers has reported that the
application of hydrostatic pressure can significantly enhance its
non-polar to polar transition temperature with a linear slope of
d$T_s$/d$P$ $\approx$ 17.54 K/GPa, reaching ca. 250 K at 6.5
GPa~\cite{paredes2018pressure}. Based on the first-principles
calculations, the enhancement of $T_s$ in \LOO~has been attributed to
the fact that pressure stabilizes the polar metallic state with a
smaller unit-cell volume than the non-polar state. The pressure effect
of increasing $T_s$ in \LOO~is different from that in BaTiO$_3$ in which
pressure reduces its $T_s$~\cite{PhysRevLett.35.1767,PhysRevLett.78.2397}.

In this work we study the pressure effect on the newly
synthesized type-II polar metal \PCOO~by monitoring the resistivity
anomaly at $T_N$ under various pressures up to 11 GPa in a cubic anvil
cell apparatus. We find that similar to \LOO, the transition temperature
$T_s=T_N$ is also increased by pressure. But interestingly, $T_N$ as a
function of pressure $P$ increases with pressure with a discontinuous
change of slope from $dT_N/dP = 6.8(8)$ K/GPa for $P < 4$ GPa to
1.8(4) K/GPa for $P > 4$ GPa. Our first-principles
calculations suggest that the change of $dT_N/dP$ can be attributed to
the vanishing of Os magnetic moment. Under pressure, Os moment is reduced
and completely suppressed above a critical value. Above the critical
pressure, the magnetic frustration is relieved and the inversion symmetry
is restored, which leads to a distinct magnetic metallic state.

\section{Experimental and computational details}
\PCOO~single crystals were grown under high-pressure and
high-temperature conditions as described
elsewhere~\cite{princep2019magnetically}.  Temperature dependences of
magnetic susceptibility and resistivity at ambient pressure were
measured on the Magnetic Properties Measurement System (MPMS-III) and
Physical Property Measurement System (PPMS-9T) from Quantum Design,
respectively. We have carried out high-pressure resistivity
measurements on \PCOO~ single crystals with standard four-probe method
by using a palm cubic anvil cell (CAC) apparatus.  Glycerol was
employed as the pressure transmitting medium and the pressure values
were determined from the pressure-loading force calibration curve at
room temperature.  
	Glycerol may produce some non-hydrostatic stress~\cite{klotz2009hydrostatic,errandonea2005pressure},
	but it is not expected to affect the reported results.
Details about the experimental setup can be found
elsewhere~\cite{cheng2014review}.

We perform spin polarized density functional theory (DFT) calculations using a plane
wave basis set and projector-augmented wave
method~\cite{PhysRevB.50.17953}, as implemented in the Vienna
Ab-initio Simulation Package
(VASP)~\cite{kresse1996efficiency,PhysRevB.54.11169}.  We take
into account spin-orbit coupling (SOC) in the DFT calculations and test the Hubbard $U$ effect. 
We use
PBEsol~\cite{Perdew2008}--a revised Perdew-Burke-Ernzerhof (PBE) generalized gradient
approximation for improving equilibrium properties of densely-packed
solids as the exchange correlation functional.  We start from the
experimental crystal structure~\cite{princep2019magnetically}, and
perform structural optimization under the studied pressures until each
Hellmann-Feynman force component is smaller than 10$^{-3}$ eV${\rm
  \AA^{-1}}$. We use the magnetic ordering obtained from experimental
neutron powder diffraction~\cite{princep2019magnetically} in all our
calculations, hence a very large cell with 80 atoms is used for the calculation.
The Brillouin zone integration is performed with a
Gaussian smearing of 0.05 eV over a $\Gamma$-centered \textbf{$k$}
mesh of 6 $\times$ 12 $\times$ 6. An energy cutoff of 600 eV is used
in all the calculations.  The threshold of energy convergence is
10$^{-7}$ eV. We use a higher energy cutoff (700 eV) and a denser
\textbf{$k$} mesh (8 $\times$ 14 $\times$ 8) and we do not find any
significant changes in the key results.
Our test calculations including Hubbard $U$ 
on Co and Os $d$ orbitals
in the approach of Dudarev
\etal~\cite{PhysRevB.57.1505} find that
electron correlation increases the magnetic moment on Co and Os. 
The DFT+$U$ calculations ($U_{\rm Co}$ = $U_{\rm Os}$ = 2 eV)
find that the magnetic moments of Co and Os are
2.75 and 1.53 $\mu_B$, respectively.
In particular, the magnetic moment of Os is
about 3.6 times the DFT-calculated Os moment of 0.43 $\mu_B$, 
which strongly deviates from the experimental value (experimentally Os moment $<$ 0.5 $\mu_B$).
Hence all the computational results presented in this study are calculated by spin polarized DFT calculations without Hubbard $U$.

\section{Results and discussion}

\begin{figure}[t]
\includegraphics[angle=0,width=0.7\textwidth]{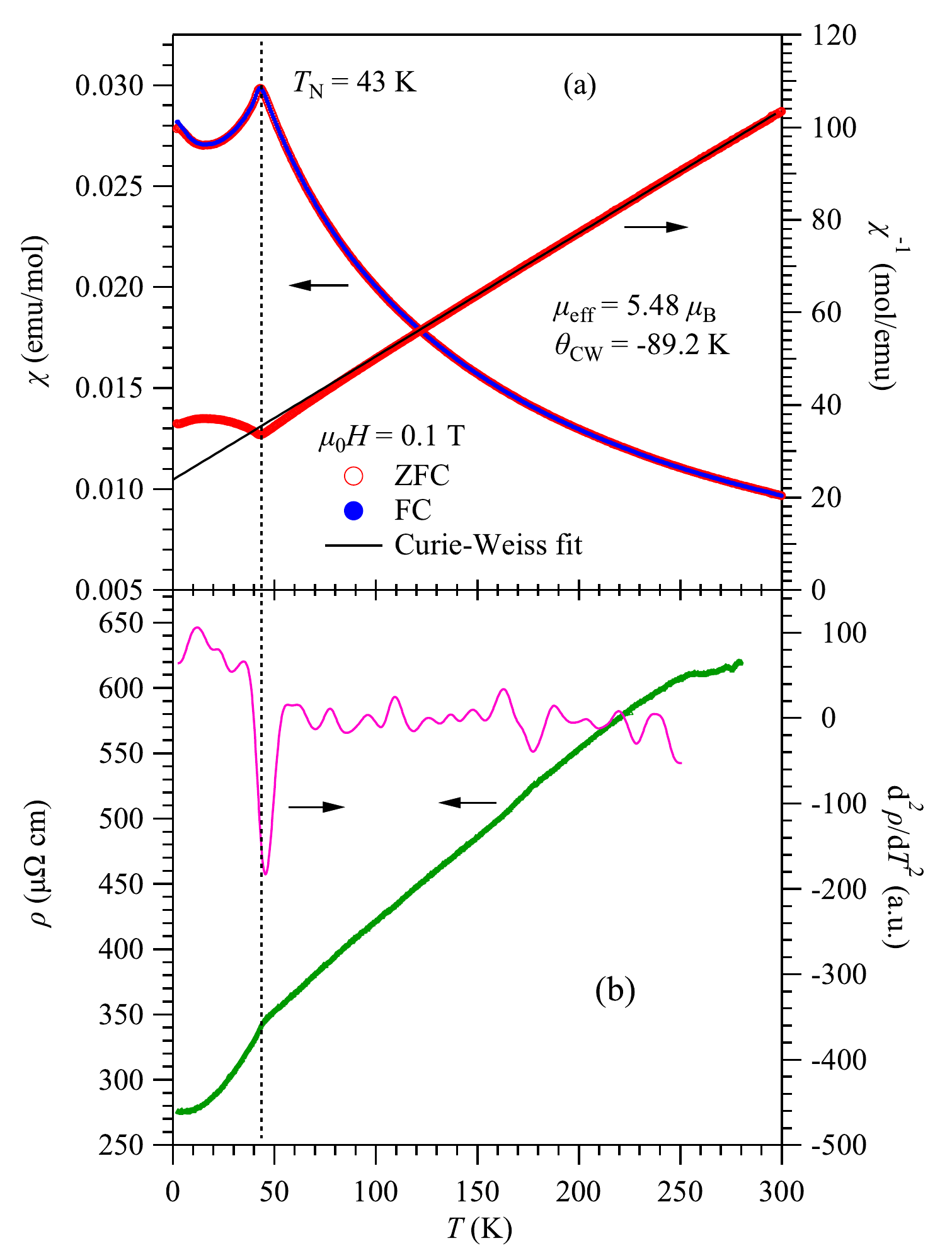}
\caption{\label{fig:fig1} Temperature dependences of (a) magnetic
  susceptibility $\chi$($T$) and its inverse $\chi^{-1}$($T$), and (b)
  resistivity $\rho$($T$) and its derivative d$\rho$/d$T$ for the
  \PCOO~ single crystal.  The antiferromagnetic transition at $T_N$ =
  43 K is marked by a vertical broken line.  }
\end{figure}

The studied Pb$_2$CoOsO$_6$ single crystal was first characterized at
ambient pressure by measuring the magnetic susceptibility $\chi$($T$)
and resistivity $\rho$($T$). The results are consistent with those
published in previous report~\cite{princep2019magnetically}. As shown
in Fig. 1, zero-field-cooled (ZFC) and field-cooled (FC) $\chi$($T$)
curves measured under ${\mu_{0}H}$ = 0.1 T are almost overlapped with
each other and both exhibit a clear cusp anomaly at the long-range
antiferromagnetic order at $T_N$ = 43 K. A Curie-Weiss fitting to
$\chi^{-1}$($T$) in the paramagnetic region above 150 K yields an
effective magnetic moment of $\mu_{\rm eff}$ = 5.48 $\mu_B$ per
formula unit and a Curie-Weiss temperature of $\theta_{\rm CW}$ =
-89.2 K, respectively. The obtained $\mu_{\rm eff}$ is larger than the
expected spin-only value of 4.8 $\mu_B$ by assuming high-spin Co(II)
with $S$ = 3/2 and Os(VI) with $S$ = 1.  This should be attributed to
the fact that high-spin Co(II) in related double perovskites always
have an effective moment higher than the spin-only value of 3.87
$\mu_B$ due to the presence of unquenched orbital
moment~\cite{morrow2013independent}. The negative $\theta_{\rm CW}$
signals the presence of dominant net antiferromagnetic interactions
that produce a moderate magnetic frustration, \ie~$|$$\theta_{\rm
  CW}$/$T_N$$|$ $\approx$ 2 in this double perovskite.  \PCOO~ is
confirmed to be metallic in the whole temperature range and its
$\rho$($T$) displays a clear inflection point at $T_N$, which can be
defined clearly from the minimum of d$^2\rho$/d$T^2$ as shown in
Fig. 1(b).  According to Ref.~\cite{princep2019magnetically}, the
antiferromagnetic ordering involving both Co and Os sublattices will
remove the inversion symmetry due to magnetic frustration, and
relax the structure into a polar structure ($P_c$). Therefore, we
can track the pressure dependence of the coupled
antiferromagnetic/structural transition by monitoring the resistivity
anomaly at $T_N$ under high pressures.

\begin{figure}[t]
\includegraphics[angle=0,width=0.8\textwidth]{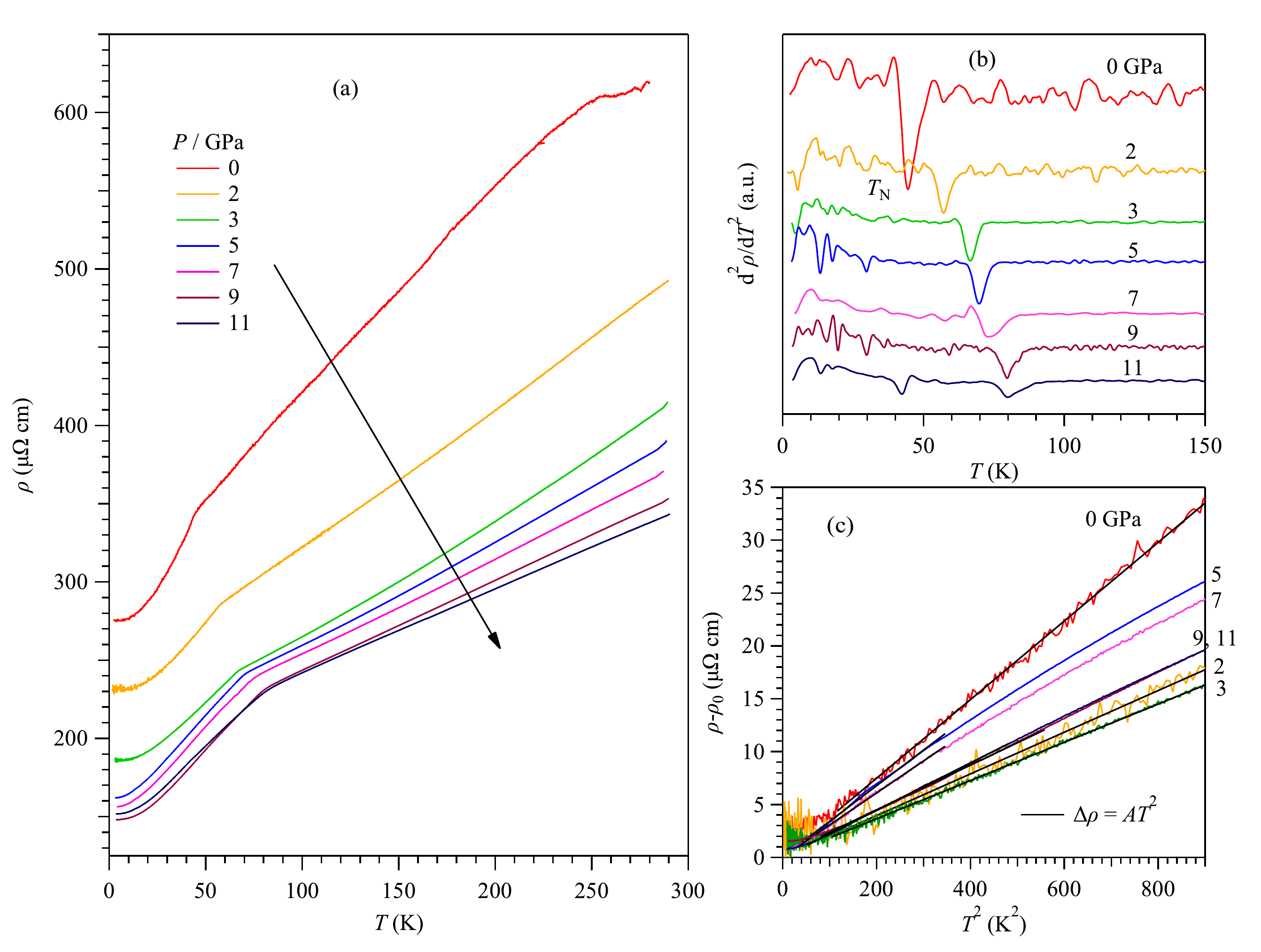}
\caption{\label{fig:fig2} Temperature dependences of (a) resistivity
  $\rho$($T$) and (b) its second derivative d$^2\rho$/d$T^2$ under
  various pressures up to 11 GPa for the \PCOO~ single crystal.  (c) A
  plot of ${\delta \rho}$ $\equiv$ ($\rho$ - $\rho_0$) versus $T^2$
  for the $\rho$($T$) data at low temperatures. The solid lines are
  linear fitting curves to extract the quadratic-$T$ coefficient $A$.
}
\end{figure}

Figure 2(a) displays the $\rho$($T$) curves of \PCOO~under various
pressures up to 11 GPa. As can be seen, it remains metallic and the
$\rho$($T$) decreases gradually with increasing pressure, in line with
the general expectation that pressure broadens the electronic
bandwidth.  The resistivity anomaly at $T_N$ is clearly visible in the
investigated pressure range and moves to higher temperatures with
increasing pressure. This can be seen more clearly from the minimum of
d$^2\rho$/d$T^2$ curves in Fig. 2(b). The determined $T_N$ are plotted
in Fig. 3(a) as a function of pressure. The transition temperature is
almost doubled and reaches $\sim$ 80 K at 11 GPa.  Interestingly, it
is found that $T_N(P)$ exhibits distinct pressure coefficients,
\ie~$T_{N}$($P$) first increases with pressure in a large slope of
+6.8(8) K/GPa for $P$ $<$ 4 GPa, and then in a much reduced slope of
+1.8(4) K/GPa for $P$ $>$ 4 GPa.

\begin{figure}[t]
\includegraphics[angle=0,width=0.7\textwidth]{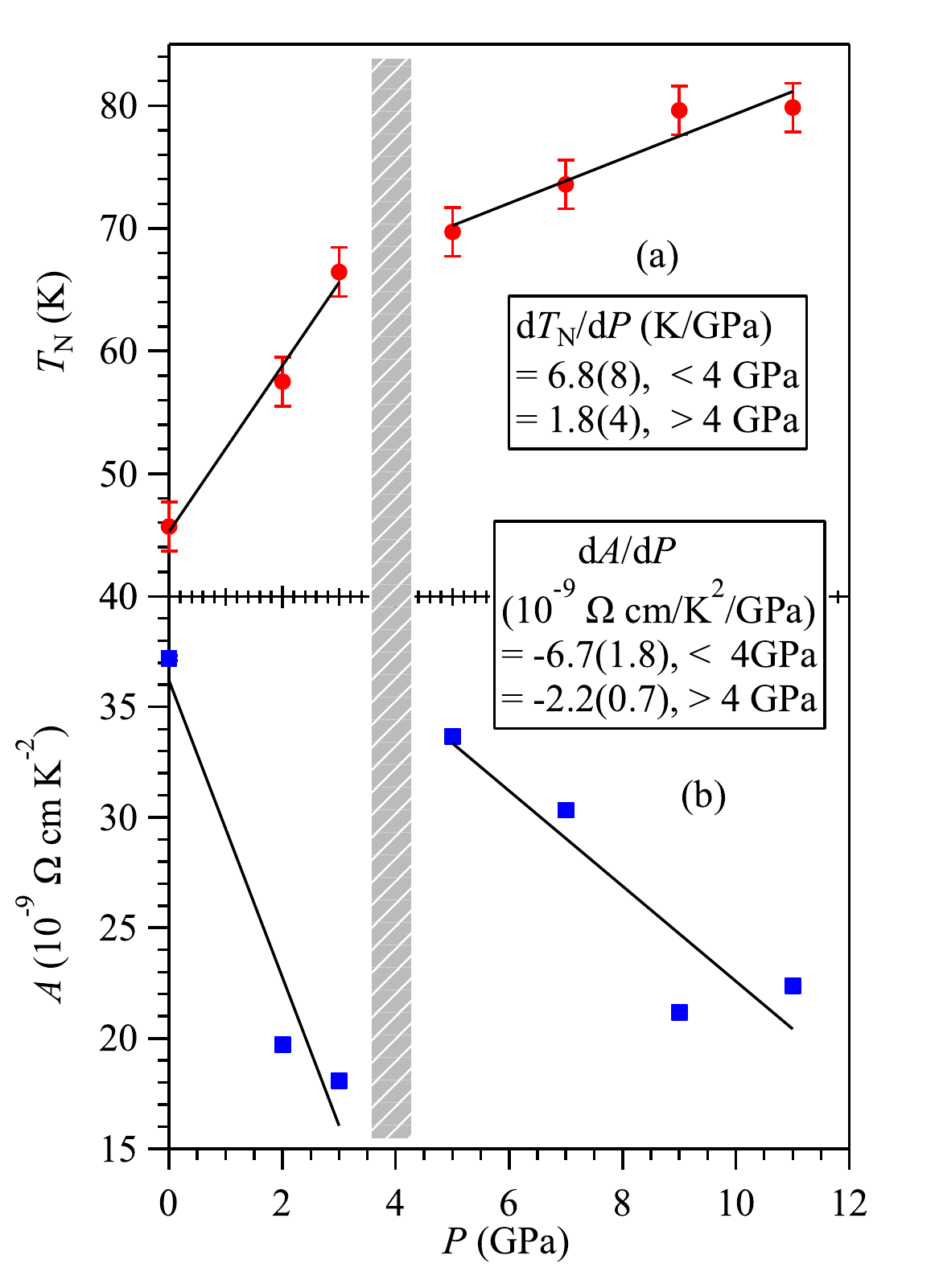}
\caption{\label{fig:fig3} Pressure dependence of (a) the
  antiferromagnetic transition temperature $T_N$ and (b) the
  quadratic-$T$ coefficient $A$ of \PCOO~single crystal. }
\end{figure}

In addition to the discontinuous slope change of $T_N$($P$), a closer
inspection of the temperature dependence of $\rho$($T$) curves at low
temperatures also evidenced a slope change for $P$ $>$ 3 GPa.  To
quantify this change, we have plotted the low-temperature $\rho$($T$)
data in the form of ($\rho$ - $\rho_0$) versus $T^2$ in Fig. 2(c),
which confirms that the Fermi-liquid behavior remains valid in the
whole pressure range. The quadratic-temperature coefficient, $A$,
determined from the linear fitting in Fig. 2(c) are shown in Fig. 3(b)
as a function of pressure.  A nonmonotonic evolution with pressure can
be clearly seen at $\sim$ 4 GPa. Similarly, $A$($P$) also exhibits a
discontinuous jump at $\sim$ 4 GPa with distinct pressure
coefficients, \ie~d$A$/d$P$ changes from 
$ -6.7 \pm 1.8 \times 10^{-9}$ $\Omega$ cm K$^{-2}$ GPa$^{-1}$ 
at $P < 4$ GPa to
$ -2.2 \pm 0.7 \times 10^{-9}$ $\Omega$ cm K$^{-2}$ GPa$^{-1}$ 
at $P > 4$ GPa.
Since the $A$ coefficient is proportional to the effective mass
of charge carriers, the distinct values of d$A$/d$P$ imply that the
impact of pressure on the electronic structure is altered
significantly at $\sim$ 4 GPa.

From these above measurements, we can reach the conclusion that the
antiferromagnetic metallic state of \PCOO~is stabilized by pressure,
and it seems to enter a distinct state above 4 GPa as illustrated by
the different pressure coefficients of $T_N$($P$) and $A$($P$). Since
the polar state blow $T_N$ at ambient pressure is driven by the
antiferromagnetic ordering involving both Co and Os sublattices, it is
essential to understand how the magnetic state evolves under
pressure in \PCOO~in order to gain some insights on the overall
structural properties, in particular polar distortions. However,
direct measurements of long-range magnetic order and magnetic
moment (especially Os moment) is difficult, even under ambient
conditions. Therefore we perform
first-principles calculations in order to elucidate why a new
distinct antiferromagnetic metallic state may emerge in \PCOO~under pressure
and to provide detailed
electronic/magnetic/structural properties of pressurized \PCOO,
which are not easily measured in experiment.

\begin{figure}[t]
\includegraphics[angle=0,width=0.8\textwidth]{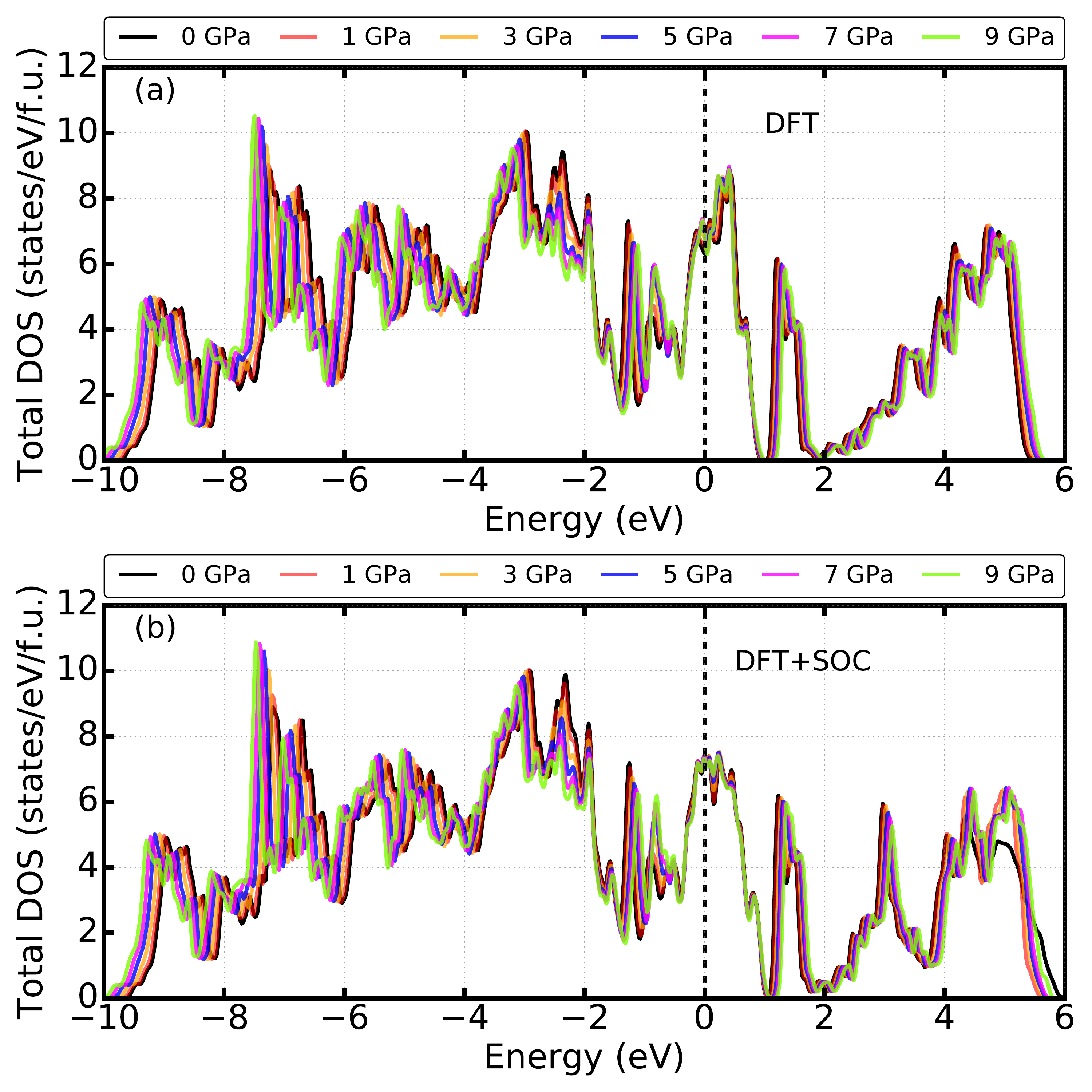}
\caption{\label{fig:TDOS} The total density of states (DOS) of \PCOO~
  under 0, 1, 3, 5, 7 and 9 GPa, obtained from (a) spin polarized DFT
  calculations and (b) DFT+SOC calculations.  Note that in panel (a),
  the spin-up and spin-down are identical due to the antiferromagnetic
  ordering, thus they are summed in the DOS. The dashed line is the
  Fermi level.  }
\end{figure}

First, we calculate the electronic structure of \PCOO~ under
pressure. The optimized crystal structure of \PCOO~at 0 K can be
  found in Table~\ref{tab:crystal} in the Appendix. Figure 4 shows
the total density of states (DOS) of \PCOO~under 0, 1, 3, 5, 7, and 9
GPa from DFT (Fig. 4(a)) and DFT+SOC (Fig. 4(b)) calculations.
Because \PCOO~has an antiferromagnetic ordering, the spin-up and
spin-down are identical in spin polarized DFT calculations, hence the
two spin-resolved DOS are summed in Fig. 4(a). In DFT+SOC
calculations, $S_z$ is no longer a good quantum number and we show the
total density of states (DOS).  We find from both DFT and DFT+SOC calculations that the
effect of pressure on the total DOS is very weak, similar to the
previous study on \LOO~\cite{paredes2018pressure}. \PCOO~remains
metallic under all the pressures in our study, consistent with the
electrical transport measurements. While standard DFT calculations
  tend to underestimate band gaps~\cite{PhysRevB.86.125202},
  we also find robust metallic properties of \PCOO~under pressure in our
  DFT+$U$ ($U_{\rm Co}$ = $U_{\rm Os}$
  = 2 eV) calculations.
By comparing the panels (a) and (b) of Fig. 4, we
find that at a given pressure, the SOC effect on the total DOS is also
very weak, especially its effect on the states close to the Fermi
level.

\begin{figure}[t]
\includegraphics[angle=0,width=0.8\textwidth]{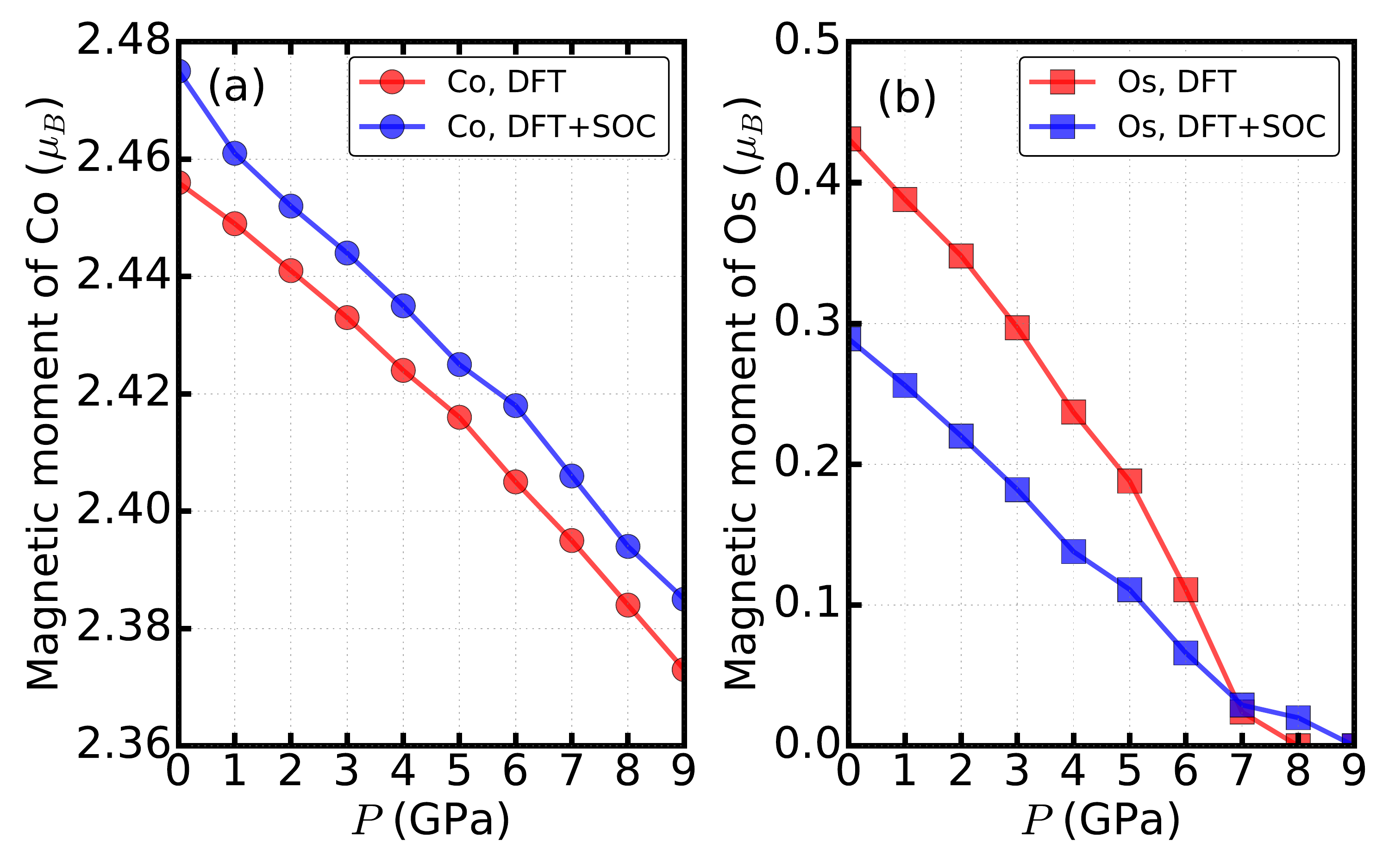}
\caption{\label{fig:magnetic-moment} Magnetic moments of (a) Co and
  (b) Os in \PCOO~ under 0$\sim$9 GPa. 
  The magnetic moments are obtained using DFT and DFT+spin-orbit coupling
    (DFT+SOC) calculations on the fully optimized crystal structures.}
  The lines are used to guide the eye.  
\end{figure}

Next, we consider the experimentally observed antiferromagnetic
ordering in \PCOO~\cite{princep2019magnetically}. Figure 5 shows the
evolution of calculated magnetic moments of Co and Os under 0$\sim$9
GPa.  DFT calculations and DFT+SOC calculations both predict that the
magnetic moments of Co and Os decrease monotonically with
increasing pressure. The SOC effect tends to decrease Os magnetic moment but
increase Co magnetic moments.
We find that from both DFT and
DFT+SOC calculations, there exists a critical pressure $P_c \sim 8$ GPa,
above which Os magnetic moment vanishes, while Co
moment only decreases slightly. This has important consequences on
the magnetic transition temperature $T_N$. When the pressure is below the
critical pressure, the N\'{e}el temperature $T_N$ of \PCOO~is determined
by the two-coupled magnetic sublattices (\ie, Co and Os). However,
above the critical pressure, $T_N$ is solely determined by the
magnetic sublattice of Co.  This may explain the discontinuity of
d$T_N$/d$P$ under pressure in our experiments.  We note that the
critical pressure estimated from our calculations is larger than the
critical pressure observed in experiment. This may result from the
approximation of exchange-correlation functional in our
first-principles calculations. While PBEsol improves the prediction
of structural properties of solids, it may overestimate the intrinsic
exchange splitting and possibly the magnitude of Os
moment~\cite{PhysRevB.93.205110,fang2019complex}.

\begin{figure}[t]
\includegraphics[angle=0,width=0.9\textwidth]{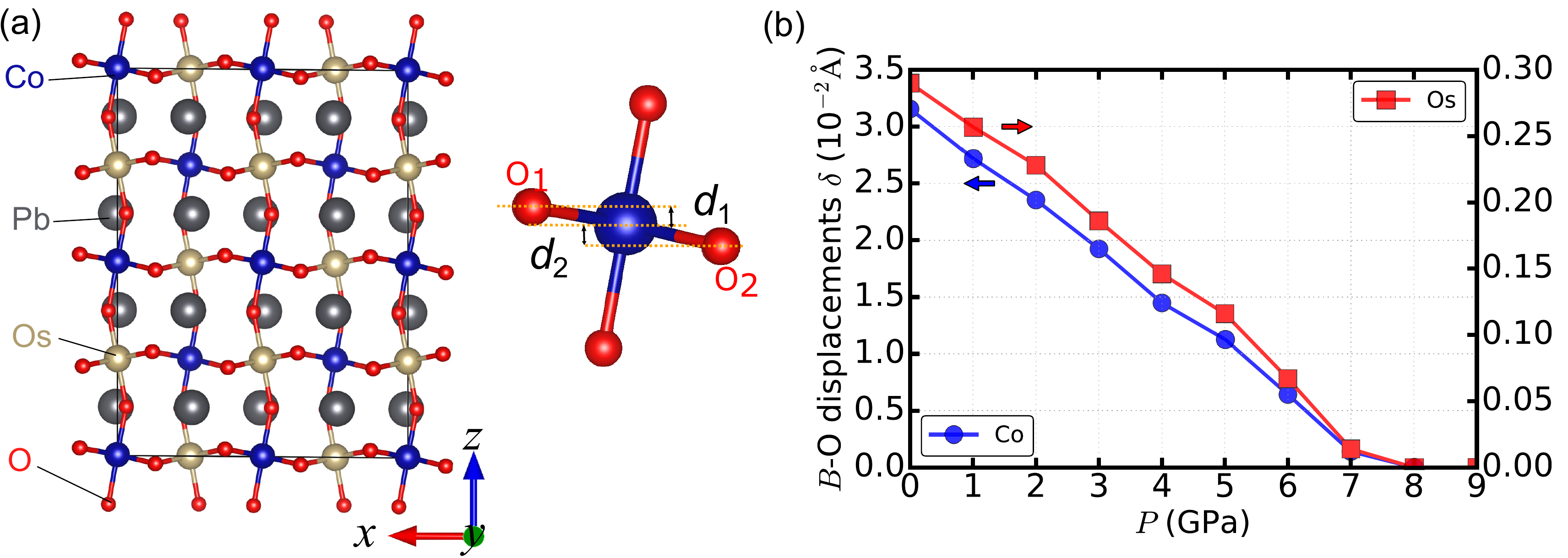}
\caption{\label{fig:displacement} (a) The optimized crystal structure
  of \PCOO~in an 80-atom cell and a metal-oxygen $B$O$_6$ octahedron ($B$ =
  Co or Os).  The polar displacement for the 
  $B$-site metal ion is defined
  as: ${\delta =\frac{1}{4}{\sum\limits_{i=1}^{4} {d_i}} }$
  where $d_{i}$ = $z_{B}$ - ${z_{O_i}}$ and the index $i$ runs over 1 to 4
  because there are four in-plane oxygen atoms in a $B$O$_6$ octahedron.
  In panel (a), only $d_1$ and $d_2$ are shown for clarity using the black
  double-headed arrows.  (b) The average Co-O and Os-O polar
  displacements $\delta$ in \PCOO~under 0$\sim$9 GPa. The lines in the
  panel (b) are used to guide the eye.  }
\end{figure}

Finally, we study the structural properties of \PCOO~under pressure.
Because each Os moment is surrounded by six Co moments and these six
Co moments form an antiferromagnetic ordering, a finite Os moment
causes magnetic frustration in this double perovskite oxide and leads
to polar distortions on both Os and Co atoms. We calculate the
polar displacements of transition metal cations (Co and Os) $\delta$ along the $z$ direction. $\delta$ is defined as
${\delta =\frac{1}{4}{\sum\limits_{i=1}^{4} {d_i}} }$, where $d_{i}$ = $z_{B}$
- ${z_{O_i}}$ and the index $i$ runs over 1 to 4 because there are
four in-plane oxygen atoms in a CoO$_6$ or OsO$_6$ octahedron (see
Fig.~\ref{fig:displacement}(a)). The polar displacement
$\delta$ almost linearly decreases
with the applied pressure up to 7 GPa, as is shown in
Fig.~\ref{fig:displacement}(b). In the range of 0 $\sim$ 7 GPa, the
Co polar displacements are substantially stronger ($\sim$ 10
times) than the Os polar displacements. Our
calculations further show that the Co-O and Os-O displacements
simultaneously vanish at 8 GPa, coincident with the complete
suppression of Os moment. This is consistent with the picture that the
inversion symmetry of \PCOO~ is broken by the magnetic frustration in
the antiferromagnetic ordering~\cite{princep2019magnetically}. With the Os moment vanishing, the
magnetic frustration is relieved and thus the inversion symmetry is
restored. In short, pressure drives \PCOO~from an antiferromagnetic
polar metal (with two types of magnetic ions) to an antiferromagnetic
centrosymmetric metal (with only one type of magnetic ion).

\section{Conclusion}

We perform transport and magnetic measurements and first-principles
calculations to study the pressure effect on the recently synthesized
type-II polar metal \PCOO. Experimentally, we monitor the resistivity
anomaly at the antiferromagnetic transition $T_N$ under various
pressures up to 11 GPa, and found a discontinuous enhancement of
$T_N$($P$) from 43 K at ambient pressure to $\sim$ 80 K at 11 GPa. The
pressure coefficient of $T_N$($P$) decreases significantly from
d$T_N$/d$P$ = 6.8(8) K/GPa for $P$ $<$ 4 GPa to 1.8(4) K/GPa for $P$
$>$ 4 GPa. Our first-principles calculations suggest that the observed
discontinuity of $dT_N/dP$ around $\sim$ 4 GPa may be attributed to the
disappearance of Os magnetic moment. Pressure substantially reduces
the Os moment and above a critical value completely suppresses the Os moment.
This relieves the magnetic frustration in the antiferromagnetic ordering
of \PCOO, which in turn decreases the
Co-O and Os-O polar displacements. Above the critical
pressure, both Co and Os atoms move to the centrosymmetric positions (i.e.
the center of CoO$_6$ and OsO$_6$ octahedra) and the inversion symmetry
is restored. This leads to a new antiferromagnetic metallic state in
pressurized \PCOO, distinct from the one under ambient pressure.

\section{Acknowledgement}
We thank Dr. Lei Hu (Tokyo Institute of Technology) 
for the valuable discussions.
This work is supported by the National Key R$\&$D Program of China
(2018YFA0305700), the National Natural Science Foundation of China
(11874400, 11834016, 11874264), the Beijing Natural Science Foundation
(Z190008), the Strategic Priority Research Program and Key Research
Program of Frontier Sciences of the Chinese Academy of Sciences
(XDB25000000 and QYZDB-SSW-SLH013) as well as the CAS
Interdisciplinary Innovation Team. Hanghui Chen is supported by the
National Natural Science Foundation of China under project number
11774236 and NYU University Research Challenge Fund. 
Kazunari Yamaura is supported by JSPS KAKENHI Grant Number JP20H05276, 
and a research grant from Nippon Sheet Glass Foundation for Materials Science and Engineering ($\#$40-37). 
NYU-HPC at
New York city, Shanghai and Abu Dhabi campuses provide
computational resources.

\appendix
\section{The DFT optimized crystal structure of \PCOO~and the discussion of Jahn-Teller distortions}
Table~\ref{tab:crystal} shows the optimized crystal structure obtained from
our DFT calculations under 0 GPa. 
The DFT-calculated crystal structures of \PCOO~under pressures up to 9 GPa are available in 
Ref.~\cite{fang_yue_wen_2020_4033535}. 

	We note that the Jahn-Teller distortions may be associated with the phase transitions
	in transition metal oxides under pressures~\cite{PhysRevLett.108.166402}, hence
we study the possible Jahn-Teller distortions in CoO$_6$ and OsO$_6$
octahedra for the optimized structures under pressures. 
We follow Ref.~\cite{PhysRevLett.101.266405} and define a Jahn-Teller
distortion parameter $\delta$ for both CoO$_6$ and OsO$_6$
octahedra: ${\delta = \frac{l - s}{2(l + s)}}$
where $l$ and $s$ are the longer and shorter $M$-O ($M$ = Os or Co)
bond lengths in the $M$O$_4$ plane.
Using this method, we find that
the Jahn-Teller distortion $\delta$ for both CoO$_6$ and OsO$_6$
octahedra are on the order of $10^{-5}$ under pressures up to 9 GPa.
This indicates that Jahn-Teller distortions in \PCOO~are very weak
both under ambient conditions and under pressures.
Therefore, we can conclude that the Jahn-Teller distortion are not correlated
to the observed magnetic transitions.

\begin{table}[!htp]
\centering
\caption{
	The cell parameters of non-centrosymmetric Pb$_2$CoOsO$_6$: $a$ = 5.6984 \AA,
	$b$ = 5.6083 \AA,  $c$ = 9.45938 \AA, $\alpha$ = 
	$\gamma$ = 90.0$^{\circ}$, 
	$\beta$ = 126.535$^{\circ}$. 
	The number of the space group of Pb$_2$CoOsO$_6$ is 7. 
	Note that we use 
the setting of standard 
	space group ${P_c}$ instead of the non-standard space group 
	${P_n}$ used in Ref.~\cite{princep2019magnetically}.
}
\label{tab:crystal}
\begin{tabular}{ccccclllll}
\hline\hline
Site & \multicolumn{1}{c}{Wyckoff Positions} & \multicolumn{1}{c}{x} & \multicolumn{1}{c}{y} & z       \\ \hline
Pb1   & 2$a$       & 0.26389      & 0.25222    & 0.25229  \\
Pb2   & 2$a$       & 0.73935      & 0.24779    & 0.75099  \\
Co    & 2$a$       & 0.00529      & 0.75035    & 0.00388  \\
Os    & 2$a$       & 0.49961      & 0.74996    & 0.50011  \\
O1    & 2$a$       & 0.19200      & 0.74949    & 0.25227  \\
O2    & 2$a$       & 0.80921      & 0.75062    & 0.74717  \\
O3    & 2$a$       & 0.29405      & 0.00817    & 0.03748  \\ 
O4    & 2$a$       & 0.70126      & 0.49262    & -0.04084 \\
O5    & 2$a$       & 0.29952      & 0.49763    & 0.03726  \\
O6    & 2$a$       & 0.69581      & 0.00158    & -0.04061 \\ \hline\hline
\end{tabular}
\end{table}

\bibliography{PCOO.bib}

\begin{thebibliography}{37}
\expandafter\ifx\csname natexlab\endcsname\relax\def\natexlab#1{#1}\fi
\expandafter\ifx\csname bibnamefont\endcsname\relax
  \def\bibnamefont#1{#1}\fi
\expandafter\ifx\csname bibfnamefont\endcsname\relax
  \def\bibfnamefont#1{#1}\fi
\expandafter\ifx\csname citenamefont\endcsname\relax
  \def\citenamefont#1{#1}\fi
\expandafter\ifx\csname url\endcsname\relax
  \def\url#1{\texttt{#1}}\fi
\expandafter\ifx\csname urlprefix\endcsname\relax\def\urlprefix{URL }\fi
\providecommand{\bibinfo}[2]{#2}
\providecommand{\eprint}[2][]{\url{#2}}

\bibitem[{\citenamefont{Eerenstein et~al.}(2006)\citenamefont{Eerenstein,
  Mathur, and Scott}}]{2006Nature-JFScott}
\bibinfo{author}{\bibfnamefont{W.}~\bibnamefont{Eerenstein}},
  \bibinfo{author}{\bibfnamefont{N.}~\bibnamefont{Mathur}}, \bibnamefont{and}
  \bibinfo{author}{\bibfnamefont{J.~F.} \bibnamefont{Scott}},
  \bibinfo{journal}{nature} \textbf{\bibinfo{volume}{442}},
  \bibinfo{pages}{759} (\bibinfo{year}{2006}).

\bibitem[{\citenamefont{Resta}(2002)}]{Resta2002JPCM}
\bibinfo{author}{\bibfnamefont{R.}~\bibnamefont{Resta}}, \bibinfo{journal}{J.
  Phys. Condens. Matter} \textbf{\bibinfo{volume}{14}}, \bibinfo{pages}{R625}
  (\bibinfo{year}{2002}).

\bibitem[{\citenamefont{Anderson and Blount}(1965)}]{PhysRevLett.14.217}
\bibinfo{author}{\bibfnamefont{P.~W.} \bibnamefont{Anderson}} \bibnamefont{and}
  \bibinfo{author}{\bibfnamefont{E.~I.} \bibnamefont{Blount}},
  \bibinfo{journal}{Phys. Rev. Lett.} \textbf{\bibinfo{volume}{14}},
  \bibinfo{pages}{217} (\bibinfo{year}{1965}).

\bibitem[{\citenamefont{Fang et~al.}(2015)\citenamefont{Fang, Ding, Tong, Zhu,
  Shen, Gong, Wan, and Duan}}]{fang2015first}
\bibinfo{author}{\bibfnamefont{Y.-W.} \bibnamefont{Fang}},
  \bibinfo{author}{\bibfnamefont{H.-C.} \bibnamefont{Ding}},
  \bibinfo{author}{\bibfnamefont{W.-Y.} \bibnamefont{Tong}},
  \bibinfo{author}{\bibfnamefont{W.-J.} \bibnamefont{Zhu}},
  \bibinfo{author}{\bibfnamefont{X.}~\bibnamefont{Shen}},
  \bibinfo{author}{\bibfnamefont{S.-J.} \bibnamefont{Gong}},
  \bibinfo{author}{\bibfnamefont{X.-G.} \bibnamefont{Wan}}, \bibnamefont{and}
  \bibinfo{author}{\bibfnamefont{C.-G.} \bibnamefont{Duan}},
  \bibinfo{journal}{Sci. Bull.} \textbf{\bibinfo{volume}{60}},
  \bibinfo{pages}{156} (\bibinfo{year}{2015}).

\bibitem[{\citenamefont{Spaldin and Ramesh}(2019)}]{spaldin2019advances}
\bibinfo{author}{\bibfnamefont{N.~A.} \bibnamefont{Spaldin}} \bibnamefont{and}
  \bibinfo{author}{\bibfnamefont{R.}~\bibnamefont{Ramesh}},
  \bibinfo{journal}{Nature materials} \textbf{\bibinfo{volume}{18}},
  \bibinfo{pages}{203} (\bibinfo{year}{2019}).

\bibitem[{\citenamefont{Dong et~al.}(2019)\citenamefont{Dong, Xiang, and
  Dagotto}}]{DongShuaiNSR2019}
\bibinfo{author}{\bibfnamefont{S.}~\bibnamefont{Dong}},
  \bibinfo{author}{\bibfnamefont{H.}~\bibnamefont{Xiang}}, \bibnamefont{and}
  \bibinfo{author}{\bibfnamefont{E.}~\bibnamefont{Dagotto}},
  \bibinfo{journal}{National Science Review} \textbf{\bibinfo{volume}{6}},
  \bibinfo{pages}{629} (\bibinfo{year}{2019}).

\bibitem[{\citenamefont{Shi et~al.}(2013)\citenamefont{Shi, Guo, Wang, Princep,
  Khalyavin, Manuel, Michiue, Sato, Tsuda, Yu et~al.}}]{Shi2013}
\bibinfo{author}{\bibfnamefont{Y.}~\bibnamefont{Shi}},
  \bibinfo{author}{\bibfnamefont{Y.}~\bibnamefont{Guo}},
  \bibinfo{author}{\bibfnamefont{X.}~\bibnamefont{Wang}},
  \bibinfo{author}{\bibfnamefont{A.~J.} \bibnamefont{Princep}},
  \bibinfo{author}{\bibfnamefont{D.}~\bibnamefont{Khalyavin}},
  \bibinfo{author}{\bibfnamefont{P.}~\bibnamefont{Manuel}},
  \bibinfo{author}{\bibfnamefont{Y.}~\bibnamefont{Michiue}},
  \bibinfo{author}{\bibfnamefont{A.}~\bibnamefont{Sato}},
  \bibinfo{author}{\bibfnamefont{K.}~\bibnamefont{Tsuda}},
  \bibinfo{author}{\bibfnamefont{S.}~\bibnamefont{Yu}}, \bibnamefont{et~al.},
  \bibinfo{journal}{Nat. Mater.} \textbf{\bibinfo{volume}{12}},
  \bibinfo{pages}{1024} (\bibinfo{year}{2013}).

\bibitem[{\citenamefont{Xiang}(2014)}]{xiang2015prb}
\bibinfo{author}{\bibfnamefont{H.~J.} \bibnamefont{Xiang}},
  \bibinfo{journal}{Phys. Rev. B} \textbf{\bibinfo{volume}{90}},
  \bibinfo{pages}{094108} (\bibinfo{year}{2014}).

\bibitem[{\citenamefont{Puggioni and Rondinelli}(2014)}]{puggioni2014designing}
\bibinfo{author}{\bibfnamefont{D.}~\bibnamefont{Puggioni}} \bibnamefont{and}
  \bibinfo{author}{\bibfnamefont{J.~M.} \bibnamefont{Rondinelli}},
  \bibinfo{journal}{Nature Communications} \textbf{\bibinfo{volume}{5}},
  \bibinfo{pages}{1} (\bibinfo{year}{2014}).

\bibitem[{\citenamefont{Fei et~al.}(2018)\citenamefont{Fei, Zhao, Palomaki,
  Sun, Miller, Zhao, Yan, Xu, and Cobden}}]{Fei2018}
\bibinfo{author}{\bibfnamefont{Z.}~\bibnamefont{Fei}},
  \bibinfo{author}{\bibfnamefont{W.}~\bibnamefont{Zhao}},
  \bibinfo{author}{\bibfnamefont{T.~A.} \bibnamefont{Palomaki}},
  \bibinfo{author}{\bibfnamefont{B.}~\bibnamefont{Sun}},
  \bibinfo{author}{\bibfnamefont{M.~K.} \bibnamefont{Miller}},
  \bibinfo{author}{\bibfnamefont{Z.}~\bibnamefont{Zhao}},
  \bibinfo{author}{\bibfnamefont{J.}~\bibnamefont{Yan}},
  \bibinfo{author}{\bibfnamefont{X.}~\bibnamefont{Xu}}, \bibnamefont{and}
  \bibinfo{author}{\bibfnamefont{D.~H.} \bibnamefont{Cobden}},
  \bibinfo{journal}{Nature} \textbf{\bibinfo{volume}{560}},
  \bibinfo{pages}{336} (\bibinfo{year}{2018}).

\bibitem[{\citenamefont{Mochizuki et~al.}(2018)\citenamefont{Mochizuki,
  Kumagai, Akamatsu, and Oba}}]{PhysRevMaterials.2.125004}
\bibinfo{author}{\bibfnamefont{Y.}~\bibnamefont{Mochizuki}},
  \bibinfo{author}{\bibfnamefont{Y.}~\bibnamefont{Kumagai}},
  \bibinfo{author}{\bibfnamefont{H.}~\bibnamefont{Akamatsu}}, \bibnamefont{and}
  \bibinfo{author}{\bibfnamefont{F.}~\bibnamefont{Oba}},
  \bibinfo{journal}{Phys. Rev. Materials} \textbf{\bibinfo{volume}{2}},
  \bibinfo{pages}{125004} (\bibinfo{year}{2018}).

\bibitem[{\citenamefont{Du et~al.}(2019)\citenamefont{Du, Lim, Zhang,
  Strohbeen, Shourov, Rodolakis, McChesney, Voyles, Fredrickson, and
  Kawasaki}}]{Du-APLMater2019}
\bibinfo{author}{\bibfnamefont{D.}~\bibnamefont{Du}},
  \bibinfo{author}{\bibfnamefont{A.}~\bibnamefont{Lim}},
  \bibinfo{author}{\bibfnamefont{C.}~\bibnamefont{Zhang}},
  \bibinfo{author}{\bibfnamefont{P.~J.} \bibnamefont{Strohbeen}},
  \bibinfo{author}{\bibfnamefont{E.~H.} \bibnamefont{Shourov}},
  \bibinfo{author}{\bibfnamefont{F.}~\bibnamefont{Rodolakis}},
  \bibinfo{author}{\bibfnamefont{J.~L.} \bibnamefont{McChesney}},
  \bibinfo{author}{\bibfnamefont{P.}~\bibnamefont{Voyles}},
  \bibinfo{author}{\bibfnamefont{D.~C.} \bibnamefont{Fredrickson}},
  \bibnamefont{and} \bibinfo{author}{\bibfnamefont{J.~K.}
  \bibnamefont{Kawasaki}}, \bibinfo{journal}{APL Materials}
  \textbf{\bibinfo{volume}{7}}, \bibinfo{pages}{121107} (\bibinfo{year}{2019}).

\bibitem[{\citenamefont{Zhang et~al.}(2019)\citenamefont{Zhang, Huang, Mei, and
  Shi}}]{PhysRevB.99.195154}
\bibinfo{author}{\bibfnamefont{H.}~\bibnamefont{Zhang}},
  \bibinfo{author}{\bibfnamefont{W.}~\bibnamefont{Huang}},
  \bibinfo{author}{\bibfnamefont{J.-W.} \bibnamefont{Mei}}, \bibnamefont{and}
  \bibinfo{author}{\bibfnamefont{X.-Q.} \bibnamefont{Shi}},
  \bibinfo{journal}{Phys. Rev. B} \textbf{\bibinfo{volume}{99}},
  \bibinfo{pages}{195154} (\bibinfo{year}{2019}).

\bibitem[{\citenamefont{Gao et~al.}(2020)\citenamefont{Gao, Fu, Yamaura, Lin,
  and Zhou}}]{PhysRevB.101.220101}
\bibinfo{author}{\bibfnamefont{J.-J.} \bibnamefont{Gao}},
  \bibinfo{author}{\bibfnamefont{S.-Y.} \bibnamefont{Fu}},
  \bibinfo{author}{\bibfnamefont{K.}~\bibnamefont{Yamaura}},
  \bibinfo{author}{\bibfnamefont{J.~F.} \bibnamefont{Lin}}, \bibnamefont{and}
  \bibinfo{author}{\bibfnamefont{J.-S.} \bibnamefont{Zhou}},
  \bibinfo{journal}{Phys. Rev. B} \textbf{\bibinfo{volume}{101}},
  \bibinfo{pages}{220101(R)} (\bibinfo{year}{2020}).

\bibitem[{\citenamefont{Volkov and Chandra}(2020)}]{PhysRevLett.124.237601}
\bibinfo{author}{\bibfnamefont{P.~A.} \bibnamefont{Volkov}} \bibnamefont{and}
  \bibinfo{author}{\bibfnamefont{P.}~\bibnamefont{Chandra}},
  \bibinfo{journal}{Phys. Rev. Lett.} \textbf{\bibinfo{volume}{124}},
  \bibinfo{pages}{237601} (\bibinfo{year}{2020}).

\bibitem[{\citenamefont{Lu et~al.}(2019)\citenamefont{Lu, Chen, Luo,
  \'I\~niguez, Bellaiche, and Xiang}}]{PhysRevLett.122.227601}
\bibinfo{author}{\bibfnamefont{J.}~\bibnamefont{Lu}},
  \bibinfo{author}{\bibfnamefont{G.}~\bibnamefont{Chen}},
  \bibinfo{author}{\bibfnamefont{W.}~\bibnamefont{Luo}},
  \bibinfo{author}{\bibfnamefont{J.}~\bibnamefont{\'I\~niguez}},
  \bibinfo{author}{\bibfnamefont{L.}~\bibnamefont{Bellaiche}},
  \bibnamefont{and} \bibinfo{author}{\bibfnamefont{H.}~\bibnamefont{Xiang}},
  \bibinfo{journal}{Phys. Rev. Lett.} \textbf{\bibinfo{volume}{122}},
  \bibinfo{pages}{227601} (\bibinfo{year}{2019}).

\bibitem[{\citenamefont{Fang and Chen}(2020)}]{fang2020design}
\bibinfo{author}{\bibfnamefont{Y.-W.} \bibnamefont{Fang}} \bibnamefont{and}
  \bibinfo{author}{\bibfnamefont{H.}~\bibnamefont{Chen}},
  \bibinfo{journal}{Communications Materials} \textbf{\bibinfo{volume}{1}},
  \bibinfo{pages}{1} (\bibinfo{year}{2020}).

\bibitem[{\citenamefont{Paredes~Aulestia
  et~al.}(2018)\citenamefont{Paredes~Aulestia, Cheung, Fang, He, Yamaura, Lai,
  Goh, and Chen}}]{paredes2018pressure}
\bibinfo{author}{\bibfnamefont{E.~I.} \bibnamefont{Paredes~Aulestia}},
  \bibinfo{author}{\bibfnamefont{Y.~W.} \bibnamefont{Cheung}},
  \bibinfo{author}{\bibfnamefont{Y.-W.} \bibnamefont{Fang}},
  \bibinfo{author}{\bibfnamefont{J.}~\bibnamefont{He}},
  \bibinfo{author}{\bibfnamefont{K.}~\bibnamefont{Yamaura}},
  \bibinfo{author}{\bibfnamefont{K.~T.} \bibnamefont{Lai}},
  \bibinfo{author}{\bibfnamefont{S.~K.} \bibnamefont{Goh}}, \bibnamefont{and}
  \bibinfo{author}{\bibfnamefont{H.}~\bibnamefont{Chen}},
  \bibinfo{journal}{Applied Physics Letters} \textbf{\bibinfo{volume}{113}},
  \bibinfo{pages}{012902} (\bibinfo{year}{2018}).

\bibitem[{\citenamefont{Princep et~al.}(2020)\citenamefont{Princep, Feng, Guo,
  Lang, Weng, Manuel, Khalyavin, Senyshyn, Rahn, Yuan
  et~al.}}]{princep2019magnetically}
\bibinfo{author}{\bibfnamefont{A.~J.} \bibnamefont{Princep}},
  \bibinfo{author}{\bibfnamefont{H.~L.} \bibnamefont{Feng}},
  \bibinfo{author}{\bibfnamefont{Y.~F.} \bibnamefont{Guo}},
  \bibinfo{author}{\bibfnamefont{F.}~\bibnamefont{Lang}},
  \bibinfo{author}{\bibfnamefont{H.~M.} \bibnamefont{Weng}},
  \bibinfo{author}{\bibfnamefont{P.}~\bibnamefont{Manuel}},
  \bibinfo{author}{\bibfnamefont{D.}~\bibnamefont{Khalyavin}},
  \bibinfo{author}{\bibfnamefont{A.}~\bibnamefont{Senyshyn}},
  \bibinfo{author}{\bibfnamefont{M.~C.} \bibnamefont{Rahn}},
  \bibinfo{author}{\bibfnamefont{Y.~H.} \bibnamefont{Yuan}},
  \bibnamefont{et~al.}, \bibinfo{journal}{Phys. Rev. B}
  \textbf{\bibinfo{volume}{102}}, \bibinfo{pages}{104410}
  (\bibinfo{year}{2020}).

\bibitem[{\citenamefont{van~den Brink and Khomskii}({2008})}]{type-II2008}
\bibinfo{author}{\bibfnamefont{J.}~\bibnamefont{van~den Brink}}
  \bibnamefont{and} \bibinfo{author}{\bibfnamefont{D.~I.}
  \bibnamefont{Khomskii}}, \bibinfo{journal}{{Journal of Physics: Condensed
  Matter}} \textbf{\bibinfo{volume}{{20}}}, \bibinfo{pages}{434217}
  (\bibinfo{year}{{2008}}).

\bibitem[{\citenamefont{Samara et~al.}(1975)\citenamefont{Samara, Sakudo, and
  Yoshimitsu}}]{PhysRevLett.35.1767}
\bibinfo{author}{\bibfnamefont{G.~A.} \bibnamefont{Samara}},
  \bibinfo{author}{\bibfnamefont{T.}~\bibnamefont{Sakudo}}, \bibnamefont{and}
  \bibinfo{author}{\bibfnamefont{K.}~\bibnamefont{Yoshimitsu}},
  \bibinfo{journal}{Phys. Rev. Lett.} \textbf{\bibinfo{volume}{35}},
  \bibinfo{pages}{1767} (\bibinfo{year}{1975}).

\bibitem[{\citenamefont{Ishidate et~al.}(1997)\citenamefont{Ishidate, Abe,
  Takahashi, and M\^ori}}]{PhysRevLett.78.2397}
\bibinfo{author}{\bibfnamefont{T.}~\bibnamefont{Ishidate}},
  \bibinfo{author}{\bibfnamefont{S.}~\bibnamefont{Abe}},
  \bibinfo{author}{\bibfnamefont{H.}~\bibnamefont{Takahashi}},
  \bibnamefont{and} \bibinfo{author}{\bibfnamefont{N.}~\bibnamefont{M\^ori}},
  \bibinfo{journal}{Phys. Rev. Lett.} \textbf{\bibinfo{volume}{78}},
  \bibinfo{pages}{2397} (\bibinfo{year}{1997}).

\bibitem[{\citenamefont{Klotz et~al.}(2009)\citenamefont{Klotz, Chervin,
  Munsch, and Le~Marchand}}]{klotz2009hydrostatic}
\bibinfo{author}{\bibfnamefont{S.}~\bibnamefont{Klotz}},
  \bibinfo{author}{\bibfnamefont{J.}~\bibnamefont{Chervin}},
  \bibinfo{author}{\bibfnamefont{P.}~\bibnamefont{Munsch}}, \bibnamefont{and}
  \bibinfo{author}{\bibfnamefont{G.}~\bibnamefont{Le~Marchand}},
  \bibinfo{journal}{Journal of Physics D: Applied Physics}
  \textbf{\bibinfo{volume}{42}}, \bibinfo{pages}{075413}
  (\bibinfo{year}{2009}).

\bibitem[{\citenamefont{Errandonea et~al.}(2005)\citenamefont{Errandonea, Meng,
  Somayazulu, and H{\"a}usermann}}]{errandonea2005pressure}
\bibinfo{author}{\bibfnamefont{D.}~\bibnamefont{Errandonea}},
  \bibinfo{author}{\bibfnamefont{Y.}~\bibnamefont{Meng}},
  \bibinfo{author}{\bibfnamefont{M.}~\bibnamefont{Somayazulu}},
  \bibnamefont{and}
  \bibinfo{author}{\bibfnamefont{D.}~\bibnamefont{H{\"a}usermann}},
  \bibinfo{journal}{Physica B: Condensed Matter}
  \textbf{\bibinfo{volume}{355}}, \bibinfo{pages}{116} (\bibinfo{year}{2005}).

\bibitem[{\citenamefont{Cheng et~al.}(2014)\citenamefont{Cheng, Matsubayashi,
  Nagasaki, Hisada, Hirayama, Hedo, Kagi, and Uwatoko}}]{cheng2014review}
\bibinfo{author}{\bibfnamefont{J.-G.} \bibnamefont{Cheng}},
  \bibinfo{author}{\bibfnamefont{K.}~\bibnamefont{Matsubayashi}},
  \bibinfo{author}{\bibfnamefont{S.}~\bibnamefont{Nagasaki}},
  \bibinfo{author}{\bibfnamefont{A.}~\bibnamefont{Hisada}},
  \bibinfo{author}{\bibfnamefont{T.}~\bibnamefont{Hirayama}},
  \bibinfo{author}{\bibfnamefont{M.}~\bibnamefont{Hedo}},
  \bibinfo{author}{\bibfnamefont{H.}~\bibnamefont{Kagi}}, \bibnamefont{and}
  \bibinfo{author}{\bibfnamefont{Y.}~\bibnamefont{Uwatoko}},
  \bibinfo{journal}{Review of Scientific Instruments}
  \textbf{\bibinfo{volume}{85}}, \bibinfo{pages}{093907}
  (\bibinfo{year}{2014}).

\bibitem[{\citenamefont{Bl\"ochl}(1994)}]{PhysRevB.50.17953}
\bibinfo{author}{\bibfnamefont{P.~E.} \bibnamefont{Bl\"ochl}},
  \bibinfo{journal}{Phys. Rev. B} \textbf{\bibinfo{volume}{50}},
  \bibinfo{pages}{17953} (\bibinfo{year}{1994}).

\bibitem[{\citenamefont{Kresse and
  Furthm{\"u}ller}(1996)}]{kresse1996efficiency}
\bibinfo{author}{\bibfnamefont{G.}~\bibnamefont{Kresse}} \bibnamefont{and}
  \bibinfo{author}{\bibfnamefont{J.}~\bibnamefont{Furthm{\"u}ller}},
  \bibinfo{journal}{Computational materials science}
  \textbf{\bibinfo{volume}{6}}, \bibinfo{pages}{15} (\bibinfo{year}{1996}).

\bibitem[{\citenamefont{Kresse and Furthm\"uller}(1996)}]{PhysRevB.54.11169}
\bibinfo{author}{\bibfnamefont{G.}~\bibnamefont{Kresse}} \bibnamefont{and}
  \bibinfo{author}{\bibfnamefont{J.}~\bibnamefont{Furthm\"uller}},
  \bibinfo{journal}{Phys. Rev. B} \textbf{\bibinfo{volume}{54}},
  \bibinfo{pages}{11169} (\bibinfo{year}{1996}).

\bibitem[{\citenamefont{Perdew et~al.}(2008)\citenamefont{Perdew, Ruzsinszky,
  Csonka, Vydrov, Scuseria, Constantin, Zhou, and Burke}}]{Perdew2008}
\bibinfo{author}{\bibfnamefont{J.~P.} \bibnamefont{Perdew}},
  \bibinfo{author}{\bibfnamefont{A.}~\bibnamefont{Ruzsinszky}},
  \bibinfo{author}{\bibfnamefont{G.~I.} \bibnamefont{Csonka}},
  \bibinfo{author}{\bibfnamefont{O.~A.} \bibnamefont{Vydrov}},
  \bibinfo{author}{\bibfnamefont{G.~E.} \bibnamefont{Scuseria}},
  \bibinfo{author}{\bibfnamefont{L.~A.} \bibnamefont{Constantin}},
  \bibinfo{author}{\bibfnamefont{X.}~\bibnamefont{Zhou}}, \bibnamefont{and}
  \bibinfo{author}{\bibfnamefont{K.}~\bibnamefont{Burke}},
  \bibinfo{journal}{Phys. Rev. Lett.} \textbf{\bibinfo{volume}{100}},
  \bibinfo{pages}{136406} (\bibinfo{year}{2008}).

\bibitem[{\citenamefont{Dudarev et~al.}(1998)\citenamefont{Dudarev, Botton,
  Savrasov, Humphreys, and Sutton}}]{PhysRevB.57.1505}
\bibinfo{author}{\bibfnamefont{S.~L.} \bibnamefont{Dudarev}},
  \bibinfo{author}{\bibfnamefont{G.~A.} \bibnamefont{Botton}},
  \bibinfo{author}{\bibfnamefont{S.~Y.} \bibnamefont{Savrasov}},
  \bibinfo{author}{\bibfnamefont{C.~J.} \bibnamefont{Humphreys}},
  \bibnamefont{and} \bibinfo{author}{\bibfnamefont{A.~P.}
  \bibnamefont{Sutton}}, \bibinfo{journal}{Phys. Rev. B}
  \textbf{\bibinfo{volume}{57}}, \bibinfo{pages}{1505} (\bibinfo{year}{1998}).

\bibitem[{\citenamefont{Morrow et~al.}(2013)\citenamefont{Morrow, Mishra,
  Restrepo, Ball, Windl, Wurmehl, Stockert, B\"uchner, and
  Woodward}}]{morrow2013independent}
\bibinfo{author}{\bibfnamefont{R.}~\bibnamefont{Morrow}},
  \bibinfo{author}{\bibfnamefont{R.}~\bibnamefont{Mishra}},
  \bibinfo{author}{\bibfnamefont{O.~D.} \bibnamefont{Restrepo}},
  \bibinfo{author}{\bibfnamefont{M.~R.} \bibnamefont{Ball}},
  \bibinfo{author}{\bibfnamefont{W.}~\bibnamefont{Windl}},
  \bibinfo{author}{\bibfnamefont{S.}~\bibnamefont{Wurmehl}},
  \bibinfo{author}{\bibfnamefont{U.}~\bibnamefont{Stockert}},
  \bibinfo{author}{\bibfnamefont{B.}~\bibnamefont{B\"uchner}},
  \bibnamefont{and} \bibinfo{author}{\bibfnamefont{P.~M.}
  \bibnamefont{Woodward}}, \bibinfo{journal}{Journal of the American Chemical
  Society} \textbf{\bibinfo{volume}{135}}, \bibinfo{pages}{18824}
  (\bibinfo{year}{2013}).

\bibitem[{\citenamefont{Ruiz-Fuertes
  et~al.}(2012{\natexlab{a}})\citenamefont{Ruiz-Fuertes, L\'opez-Moreno,
  L\'opez-Solano, Errandonea, Segura, Lacomba-Perales, Mu\~noz, Radescu,
  Rodr\'{\i}guez-Hern\'andez, Gospodinov et~al.}}]{PhysRevB.86.125202}
\bibinfo{author}{\bibfnamefont{J.}~\bibnamefont{Ruiz-Fuertes}},
  \bibinfo{author}{\bibfnamefont{S.}~\bibnamefont{L\'opez-Moreno}},
  \bibinfo{author}{\bibfnamefont{J.}~\bibnamefont{L\'opez-Solano}},
  \bibinfo{author}{\bibfnamefont{D.}~\bibnamefont{Errandonea}},
  \bibinfo{author}{\bibfnamefont{A.}~\bibnamefont{Segura}},
  \bibinfo{author}{\bibfnamefont{R.}~\bibnamefont{Lacomba-Perales}},
  \bibinfo{author}{\bibfnamefont{A.}~\bibnamefont{Mu\~noz}},
  \bibinfo{author}{\bibfnamefont{S.}~\bibnamefont{Radescu}},
  \bibinfo{author}{\bibfnamefont{P.}~\bibnamefont{Rodr\'{\i}guez-Hern\'andez}},
  \bibinfo{author}{\bibfnamefont{M.}~\bibnamefont{Gospodinov}},
  \bibnamefont{et~al.}, \bibinfo{journal}{Phys. Rev. B}
  \textbf{\bibinfo{volume}{86}}, \bibinfo{pages}{125202}
  (\bibinfo{year}{2012}{\natexlab{a}}).

\bibitem[{\citenamefont{Chen and Millis}(2016)}]{PhysRevB.93.205110}
\bibinfo{author}{\bibfnamefont{H.}~\bibnamefont{Chen}} \bibnamefont{and}
  \bibinfo{author}{\bibfnamefont{A.~J.} \bibnamefont{Millis}},
  \bibinfo{journal}{Phys. Rev. B} \textbf{\bibinfo{volume}{93}},
  \bibinfo{pages}{205110} (\bibinfo{year}{2016}).

\bibitem[{\citenamefont{Fang et~al.}(2019)\citenamefont{Fang, Yang, and
  Chen}}]{fang2019complex}
\bibinfo{author}{\bibfnamefont{Y.-W.} \bibnamefont{Fang}},
  \bibinfo{author}{\bibfnamefont{R.}~\bibnamefont{Yang}}, \bibnamefont{and}
  \bibinfo{author}{\bibfnamefont{H.}~\bibnamefont{Chen}},
  \bibinfo{journal}{Journal of Physics: Condensed Matter}
  \textbf{\bibinfo{volume}{31}}, \bibinfo{pages}{445803}
  (\bibinfo{year}{2019}).

\bibitem[{\citenamefont{Fang and Chen}()}]{fang_yue_wen_2020_4033535}
\bibinfo{author}{\bibfnamefont{Y.-W.} \bibnamefont{Fang}} \bibnamefont{and}
  \bibinfo{author}{\bibfnamefont{H.}~\bibnamefont{Chen}},
  \emph{\bibinfo{title}{{Dataset for coupled magnetic and structural phase
  transitions in antiferromagnetic polar metal Pb$_2$CoOsO$_6$ under pressure
  }}}, \bibinfo{howpublished}{\url{https://doi.org/10.5281/zenodo.4033535}}.

\bibitem[{\citenamefont{Ruiz-Fuertes
  et~al.}(2012{\natexlab{b}})\citenamefont{Ruiz-Fuertes, Segura,
  Rodr\'{\i}guez, Errandonea, and Sanz-Ortiz}}]{PhysRevLett.108.166402}
\bibinfo{author}{\bibfnamefont{J.}~\bibnamefont{Ruiz-Fuertes}},
  \bibinfo{author}{\bibfnamefont{A.}~\bibnamefont{Segura}},
  \bibinfo{author}{\bibfnamefont{F.}~\bibnamefont{Rodr\'{\i}guez}},
  \bibinfo{author}{\bibfnamefont{D.}~\bibnamefont{Errandonea}},
  \bibnamefont{and} \bibinfo{author}{\bibfnamefont{M.~N.}
  \bibnamefont{Sanz-Ortiz}}, \bibinfo{journal}{Phys. Rev. Lett.}
  \textbf{\bibinfo{volume}{108}}, \bibinfo{pages}{166402}
  (\bibinfo{year}{2012}{\natexlab{b}}).

\bibitem[{\citenamefont{Pavarini et~al.}(2008)\citenamefont{Pavarini, Koch, and
  Lichtenstein}}]{PhysRevLett.101.266405}
\bibinfo{author}{\bibfnamefont{E.}~\bibnamefont{Pavarini}},
  \bibinfo{author}{\bibfnamefont{E.}~\bibnamefont{Koch}}, \bibnamefont{and}
  \bibinfo{author}{\bibfnamefont{A.~I.} \bibnamefont{Lichtenstein}},
  \bibinfo{journal}{Phys. Rev. Lett.} \textbf{\bibinfo{volume}{101}},
  \bibinfo{pages}{266405} (\bibinfo{year}{2008}).

\end{thebibliography}

\end{document}